\documentclass[12pt]{article}
\usepackage{amsfonts}
\usepackage{amssymb}
\usepackage{graphics,psboxit,amsmath}
\usepackage{subfigure}
\usepackage{graphicx}
\usepackage{verbatim}
\usepackage{epic}

\def\hybrid{\topmargin 0pt \oddsidemargin 0pt 
        \headheight 0pt \headsep 0pt
        \textwidth 16,5cm 
        \textheight 23cm 
        \marginparwidth .875in
        \parskip 5pt plus 1pt \jot = 1.5ex}

\hybrid

\catcode`\@=11

\def\marginnote#1{}
\newcount\hour
\newcount\minute
\newtoks\amorpm
\hour=\time\divide\hour by60
\minute=\time{\multiply\hour by60 \global\advance\minute by-\hour}
\edef\standardtime{{\ifnum\hour<12 \global\amorpm={am}%
        \else\global\amorpm={pm}\advance\hour by-12 \fi
        \ifnum\hour=0 \hour=12 \fi
        \number\hour:\ifnum\minute<10 0\fi\number\minute\the\amorpm}}
\edef\militarytime{\number\hour:\ifnum\minute<10 0\fi\number\minute}

\def\draftlabel#1{{\@bsphack\if@filesw {\let\thepage\relax
   \xdef\@gtempa{\write\@auxout{\string
      \newlabel{#1}{{\@currentlabel}{\thepage}}}}}\@gtempa
   \if@nobreak \ifvmode\nobreak\fi\fi\fi\@esphack}
        \gdef\@eqnlabel{#1}}
\def\@eqnlabel{}
\def\@vacuum{}
\def\draftmarginnote#1{\marginpar{\raggedright\scriptsize\tt#1}}

\def\draft{\oddsidemargin -.5truein
        \def\@oddfoot{\sl preliminary draft \hfil
        \rm\thepage\hfil\sl\today\quad\militarytime}
        \let\@evenfoot\@oddfoot \overfullrule 3pt
        \let\label=\draftlabel
        \let\marginnote=\draftmarginnote
   \def\@eqnnum{(\theequation)\rlap{\kern\marginparsep\tt\@eqnlabel}%
\global\let\@eqnlabel\@vacuum} }

\def\draft2{
        \def\@oddfoot{\sl preliminary draft \hfil
        \rm\thepage\hfil\sl\today\quad\militarytime}
        \let\@evenfoot\@oddfoot \overfullrule 3pt
        \let\label=\draftlabel
        \let\marginnote=\draftmarginnote
   \def\@eqnnum{(\theequation)\rlap{\kern\marginparsep\tt\@eqnlabel}%
\global\let\@eqnlabel\@vacuum} }

\def\preprint{\twocolumn\sloppy\flushbottom\parindent 2em
        \leftmargini 2em\leftmarginv .5em\leftmarginvi .5em
        \oddsidemargin -.5in \evensidemargin -.5in
        \columnsep .4in \footheight 0pt
        \textwidth 10.in \topmargin -.4in
        \headheight 12pt \topskip .4in
        \textheight 6.9in \footskip 0pt
        \def\@oddhead{\thepage\hfil\addtocounter{page}{1}\thepage}
        \let\@evenhead\@oddhead \def\@oddfoot{} \def\@evenfoot{} }

\def\numberbysection{\@addtoreset{equation}{section}
        \def\theequation{\thesection.\arabic{equation}}}

\def\underline#1{\relax\ifmmode\@@underline#1\else
        $\@@underline{\hbox{#1}}$\relax\fi}

\def\titlepage{\@restonecolfalse\if@twocolumn\@restonecoltrue\onecolumn
     \else \newpage \fi \thispagestyle{empty}\c@page\z@
        \def\thefootnote{\fnsymbol{footnote}} }

\def\endtitlepage{\if@restonecol\twocolumn \else \newpage \fi
        \def\thefootnote{\arabic{footnote}}
        \setcounter{footnote}{0}} 

\catcode`@=12
\relax

\def\figcap{\section*{Figure Captions\markboth
        {FIGURECAPTIONS}{FIGURECAPTIONS}}\list
        {Figure \arabic{enumi}:\hfill}{\settowidth\labelwidth{Figure
999:}
        \leftmargin\labelwidth
        \advance\leftmargin\labelsep\usecounter{enumi}}}
 \relax
\def\tablecap{\section*{Table Captions\markboth
        {TABLECAPTIONS}{TABLECAPTIONS}}\list
        {Table \arabic{enumi}:\hfill}{\settowidth\labelwidth{Table
999:}
        \leftmargin\labelwidth
        \advance\leftmargin\labelsep\usecounter{enumi}}}
 \relax
\def\reflist{\section*{References\markboth
        {REFLIST}{REFLIST}}\list
        {[\arabic{enumi}]\hfill}{\settowidth\labelwidth{[999]}
        \leftmargin\labelwidth
        \advance\leftmargin\labelsep\usecounter{enumi}}}
 \relax
\makeatletter
\newcounter{pubctr}
\def\publist{\@ifnextchar[{\@publist}{\@@publist}}
\def\@publist[#1]{\list
        {[\arabic{pubctr}]\hfill}{\settowidth\labelwidth{[999]}
        \leftmargin\labelwidth
        \advance\leftmargin\labelsep
        \@nmbrlisttrue\def\@listctr{pubctr}
        \setcounter{pubctr}{#1}\addtocounter{pubctr}{-1}}}
\def\@@publist{\list
        {[\arabic{pubctr}]\hfill}{\settowidth\labelwidth{[999]}
        \leftmargin\labelwidth
        \advance\leftmargin\labelsep
        \@nmbrlisttrue\def\@listctr{pubctr}}}
 \relax
\makeatother

\def\ba{\begin{equation}}
\def\ea{\end{equation}}

\def\del{\partial}

\def\r{\rho}

\def\d{\delta}
\def\D{\Delta}

\def\th{\theta}

\def\m{\mu}

\def\Om{\Omega}

\def\s{\sigma}

\def\cL{{\cal L}}

\def\cN{{\cal N}}

\def\no{\noindent}

\def\qq{\qquad}

\def\IR{\relax{\rm I\kern-.18em R}}

\def \ha {{1\over 2}}

\def \ov {\over}

\begin{document}


\renewcommand{\theequation}{\thesection.\arabic{equation}}
\csname @addtoreset\endcsname{equation}{section}

\newcommand{\eqn}[1]{(\ref{#1})}
\newcommand{\be}{\begin{eqnarray}}
\newcommand{\ee}{\end{eqnarray}}
\newcommand{\non}{\nonumber}
\begin{titlepage}
\strut\hfill
\vskip 1.3cm
\begin{center}

\vskip -2 cm


\vskip 2 cm

{\large \bf NS5-branes, holography and CFT deformations}
\footnote{{\tt Proceedings contribution to the {\it 9th Hellenic School on
Elementary Particle Physics and Gravity}, Corfu, Greece, September 2009. Based on a talk given by K.S.\hfill}}

\vskip 0.5in

{\bf A. Fotopoulos}$^{1}$, {\bf P.M. Petropoulos}$^{2}$, {\bf N. Prezas}$^{3}$
and {\bf K. Sfetsos}$^{4}$

\vskip 0.1in

${}^1$Dipartimento di Fisica Teorica dell'Universit\`a di Torino
and INFN,\\Sezione di Torino,
via P. Giuria 1, 10125 Torino, Italy

\vskip 0.1in

${}^2$Centre de Physique Th{\'e}orique, Ecole Polytechnique, CNRS--UMR 7644,\\
91128 Palaiseau Cedex, France.

\vskip 0.1in

${}^3$Albert Einstein Center for Fundamental Physics
Institute for Theoretical Physics,\\
University of Bern, Sidlerstrasse 5, CH-3012 Bern, Switzerland.

\vskip 0.1in

${}^4$Department of Engineering Sciences, University of Patras,\\
26110 Patras, Greece

\vskip .1in

\vskip .15in

\end{center}

\vskip .4in

\centerline{\bf Abstract}

\no
 A few supergravity solutions representing
configurations of NS5-branes admit exact conformal
field theory (CFT) description. Deformations of these solutions should be
described by exactly marginal operators of the corresponding theories. We
briefly review the essentials of these constructions and present, as
a new case, the operators responsible for turning on angular
momentum.

\vfill
\no


\end{titlepage}
\vfill
\eject



\def\baselinestretch{1.0}
\baselineskip 15 pt
\no

\section{Introduction: NS5-branes and basic CFTs}

Branes were involved in the most important developments in string theory,
from a deeper understanding of the theory itself,
to black hole physics and the AdS/CFT correspondence.
In string theory it is possible, in some rare occasions, to go beyond the supergravity approximation
and obtain an exact CFT description of the solutions of interest.
In particular,
among the plethora of the various brane configurations a tiny subset
involving exclusively NS5 or NS1 and NS5-branes admit, under certain
circumstances, such a description.

\no
Consider $k$ parallel NS5-branes, spread out in the transverse $R^4$ with density $\r(\bf x)$
normalized to unity. The half-supersymmetric preserving solitonic solution has metric
given by \cite{duff}
\be
ds^2 =  \underbrace{\eta_{\mu \nu} dy^\mu dy^\nu}_{ \rm 6-dim\ flat} \quad
+  \underbrace{H ({\bf x} ) \ \delta_{ij}
dx^i dx^j}_{\rm 4\!-\!dim\ non-trivial\ part}\ ,\qq
H({\bf x}) = 1+ \alpha' k \int_{R^4} d^4 x'
\frac{\rho({\bf x'})}{|{\bf x}-{\bf x'}|^2}\ ,
\ee
where $H$ is a harmonic function in $R^4$. It is supported by
a NS three-form and a dilaton
\be
H_{ijk}=\epsilon_{ijk}^{\phantom{ijk}l} \partial_l H\ ,\qq e^{2 (\Phi-\Phi_0) }=H \ .
\ee
Indices are raised and lowered with the flat metric of
$R^4$ and $e^{\Phi_0}=g_s$ is the asymptotic string coupling.

\no
\underline{Branes at a point \cite{CHS}:}
In this case and in the near horizon limit (1 dropped in $H$) the background is
\be
ds^2 = \eta_{\mu \nu} dy^\mu dy^\nu + d\phi^2 + 2 k d\Om_3^2\ ,\qquad H= 2 {\rm Vol}_{S^3}\ ,\qquad
\Phi = -{q\ov 2}\ \phi\ ,
\label{jk9}
\ee
where, for the radial distance, we let
$r = \sqrt{2 k} e^{\Phi_0 + \phi/\sqrt{2 k}}$ ($\alpha'=2$) and also $q=\sqrt{2\ov k}$.
This will be the geometry of every localized NS5-brane distribution far from it.
The above background corresponds to the exact CFT with $\cN = 4$ worldsheet supersymmetry
\be
R^{5,1} \times R_\phi \times SU(2)_k\ ,
\label{linee}
\ee
where $R_\phi$ is the  linear dilaton factor with background charge $q$.

\no
\underline{Branes on a circle \cite{sfecircle}:}
When the branes are located at the corners of a canonical $k$-polygone,
the exact $\cN = 4$ supersymmetric CFT, in the near horizon limit, is
\be
R^{5,1} \times {SU(2)_k\over U(1)} \times {SL(2,R)_{-k}\over U(1)}\ ,
\label{dhi1}
\ee
orbifolded under a discrete $Z_k$ symmetry whose geometric origin
is the rotation symmetry of the canonical $k$-polygone. In the continuum
limit the distribution is over the circumference of a circle.

\no
The question we have addressed in a series of papers is how, mainly, world-volume
preserving deformations of the above solutions,
can be described by exact operators of the corresponding CFTs.
This approach was initiated for the CFT (\ref{dhi1}) in
\cite{petrosfe1}, by describing the bosonic content of the perturbation that deforms the circular
distribution into an ellipsoidal one and
perfected in its full technical and conceptual details in \cite{FPPS}
by including the fermionic part of the deformation as well as arbitrary deformations.
In \cite{PreSfe} all possible deformations of the solution (\ref{jk9})
starting form the CFT in (\ref{linee}) were classified.
We refer, to these works for the details of the construction that we present below in section 2.
Finally, we have considered deformations based on the
CFT $SL(2,R) \times SU(2) \times U(1)^4$ arising in the near-horizon limit
of a system of NS5-branes wrapped on a 4-torus and NS1-branes smeared
completely on the 4-torus \cite{PPS}.

\section{Deforming the solutions with exact CFT operators}

In the NS5's worldvolume there is the so called little string theory (LST) \cite{Aharony:1998ub}. It is obtained in the
$g_s\to 0$ limit and therefore is non-gravitational.
It can be considered as a UV completion of a six-dimensional gauge theory with 16 supercharges
and $g_{\rm YM}^2\sim \alpha'$, where asymptotically linear dilaton backgrounds
provide holographic duals.
We will address the following issues: (i) what are the operators of the exact CFTs
responsible for deforming the supergravity solutions mentioned above and (ii) what is their holographic interpretation
in terms of the LST.
We focus on cases where the deformation affects only the distribution of the
NS5-branes in the transverse space,
thus preserving Lorentz symmetry in their worldvolume.

\no
\underline{Spreading the branes out of the point:} We present a brief summary of the analysis
of \cite{PreSfe} where the interested reader can find all details.
Consider the operators $\widetilde {\rm  tr} (X^{i_1}X^{i_2}\cdots X^{i_{2j+2}})$, with $i_\ell=1,2,3,4$\ ,
where the $SO(4)$ scalars $X^i$ are $k\times k$ traceless matrices,
in the adjoint of $SU(k)$.
In order for the LST operators to be
in a short multiplet of spacetime supersymmetry, only the
traceless and symmetric components in the indices $i_1,\ldots,i_{2j+2}$ should be kept.
The tilde on the trace means that we should not consider the standard single trace,
but its combination with multi-traces. This will not play any
r\^ole for our considerations.
Geometrically,
the eigenvalues of the $X^i$'s parametrize the transverse positions of the NS5's.

\no
The correspondence
should involve the primaries $\Phi^{\rm su}_{j;m,\bar m}$
of the bosonic $SU(2)_{k-2}$ WZW model in (\ref{linee}). These are realized,
semiclassically, in terms of the Euler angles ${\th, \phi_1, \phi_2}$ parametrizing the $SU(2)$ group element.
In addition, it should involve the corresponding fermions $\psi_\pm$ and $\psi_3$ in the adjoint of $SU(2)$.
Finally, one may use the boson $\phi$ and the corresponding fermion $\psi_\phi$, as well as
the corresponding vertex operator $e^{-q a_j \phi}$ of the linear dilaton factor ${R}_\phi$.
The precise holographic correspondence is \cite{Aharony:2003vk,Aharony:2004xn}
\be
\overbrace{\widetilde {\rm  tr} (X^{i_1}X^{i_2}\cdots X^{i_{2j+2}})}^{\rm LST }
 \qq  \Longleftrightarrow \qq
{\cal V}_{j;m,\bar m}=
\overbrace{(\psi\bar \psi \Phi^{\rm su}_j)_{j+1;m,\bar m} e^{-q a_j \phi}}^{\rm CFT } \ .
\label{holodic}
\ee
This is justified as follows:
First notice that the scalars
transform in the $(\ha,\ha)$ representation of $SU(2)_L\times SU(2)_R$, so that the left
hand side of this correspondence has spin $j+1$. On the right hand side the operator should have the same spin.
Using CFT operators it reads (we suppress the antiholomorphic indices)
\be
{\cal V}_{j;m}=
\left(\m_3 \psi^3 \Phi_{j,m_3}^{\rm su} +  \m_+ \psi^+ \Phi_{j,m_+}^{\rm su}
+  \m_- \psi^- \Phi_{j,m_-}^{\rm su}\right)e^{-qa_j  \phi}\ .
\ee
In order to have spin $j+1$, the constants $\m_{3,\pm}$ are chosen to be the
Clebsch--Gordan coefficients arising from composing a spin $j$ state (from the bosonic $SU(2)_{k-2}$)
with a spin 1 state (where the fermions belong) and $m_3=m$, $m_\pm =m\mp 1$.
In addition, the $U(1)$'s symmetries in $R^4$ (rotations on the planes $x^1\!-\!x^2$ and $x^3\!-\!x^4$)
are associated with the quantum numbers $m$ and $\bar m$.
Finally, we note that from standard CFT unitarity arguments for the $SU(2)_{k-2}$ current algebra,
the spin $j$ is bounded so that $j=0,\ha,\dots ,(k-2)/2$. Hence,
the number of matrices on the left hand side is $k$, the same as the dimension of the matrices $X^i$.

\no
The coefficient $a_j$ of the dilaton  vertex operator on the right
must be either $a_j=j+1$ or $a_j=-j$ in order for the actual deforming operators,
constructed in (\ref{pertti}) below, to be marginal.
In the first case the operator
is normalizable and
corresponds to a situation where the
dual  LST operator acquires a vacuum expectation value (vev). In the second case it corresponds to a
non-normalizable deformation of the theory that triggers
a perturbation of the LST with the dual operator.
The $\cN =1$ worldsheet preserving perturbation is
\be
\sum_{j=0}^{(k-2)/2}\sum_{m,\bar m=-(j+1)}^{j+1}
\left(\lambda_{j;m,\bar m} G_{-\frac{1}{2}}\bar G_{-\frac{1}{2}}
 (\psi\bar \psi \Phi_j^{{\rm su}})_{j+1;m,\bar m}
e^{-qa_j\phi} +{\rm c.c.} \right)\ ,
\label{pertti}
\ee
where $G$ is the ${\cal N}=1$ supercurrent, realized in terms of $SU(2)_{k-2}$ currents, $\phi$ and
the fermions \cite{Kounnas}. Giving vev's to the $X^i$'s, dictating the density $\r(\textbf{x})$,
determines the coupling as
\be
\lambda_{j;m,\bar m}\sim  \widetilde {\rm tr} (X^{i_1}X^{i_2}\cdots X^{i_{2j+2}})\ .
\ee
The perturbation produces a purely bosonic term, and two fermionic terms, one quadratic and one
quartic. It should be cast in the form of
a supersymmetric $\sigma$-model (see, for instance, \cite{Howe}).

\no
To describe a deformation of the brane's position and not a flow
to a different theory, the perturbation
should preserve $\cN=4$ supersymmetry (and consequently spacetime supersymmetry),
but only in the normalizable
branch.\footnote{There are
non-normalizable operators
preserving ${\cal N}=4$ worldsheet supersymmetry (and spacetime),
 that do not correspond to an LST deformation, but
they perturb instead away from the NS5-brane horizon \cite{PreSfe}.}
Using the operators of $\cN=4$ realized in terms of the $SU(2)_{k-2}$ currents,
$\phi$ and the fermions we find that indeed
$a_j=j+1$. 
In addition,
not all values of $m$, for fixed $j$, are allowed, but
\be
m=\pm(j+1): \qq {\cal V}_{j;\pm(j+1)}=\psi^+ \Phi^{\rm su}_{j,\pm j} e^{-q(j+1)\phi}\ ,
\ee
which for the upper (lower) sign are chiral (anti-chiral) primaries and have $h=\pm q/2$.
A quite important exception is the operator
$\psi_3 e^{-q \phi}$, which is primary, but not chiral (or antichiral).
Assuming that superconformal invariance remains unbroken
the perturbation (\ref{pertti}) is exactly marginal.

\no
Geometrically, one can summarize
the distinct cases arising in fig.~1.

\vskip 0 cm
\begin{figure}[htp]
\begin{center}
\includegraphics[height= 6 cm,angle=0]{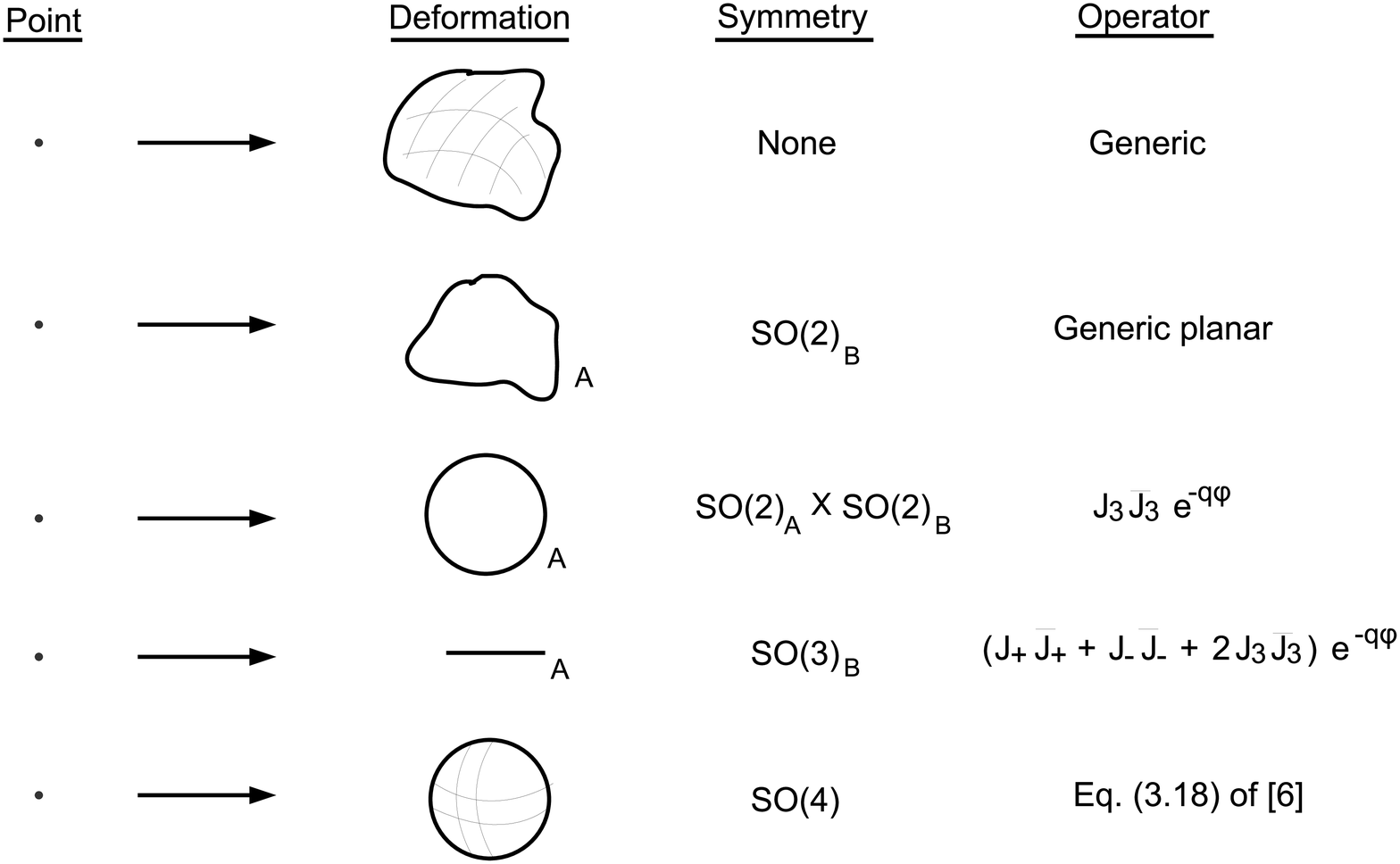}
\end{center}
\vskip -0.3 cm
\caption{NS5-brane configurations in the continuous limit, their symmetries and the bosonic part
of the operators
driving the perturbation: (i) generic distribution in ${R}^4$ with
no symmetry, (ii) generic planar deformation, (iii) circle
(realizing a dilaton domain wall \cite{PreSfe}), (iv) line segment , (v) 3-sphere.
A and B label the planes where the branes are distributed (for planar distributions)
and the one orthogonal to that, respectively. The
 vertex operator $e^{a\phi}$ has conformal dimension $\D=-\ha a (a+q)$.}
\end{figure}

\no
\underline{Deformation of a finite size circle:} In this case, depicted in  fig.~2, the appropriate
CFT is (\ref{dhi1}).
Now the bosonic part of the perturbation is in terms of compact parafermions
dressed with primaries of the non-compact part of the theory \cite{petrosfe1,FPPS}. The
fermions enter through the bosons we use for their bosonization. The important new feature is that
for finite $k$, there are no purely bosonic or fermionic terms. The clear semiclassical $\s$-model picture appears only
in the limit $k\gg 1$. All the details can be found in \cite{FPPS}.
\vskip 0cm
\begin{figure}[htp]
\begin{center}
\includegraphics[height= 1.3 cm,angle=0]{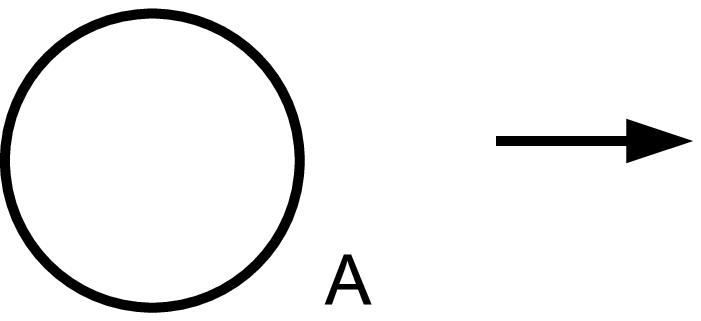}\hskip 1 cm
\includegraphics[height= 1.3 cm,angle=0]{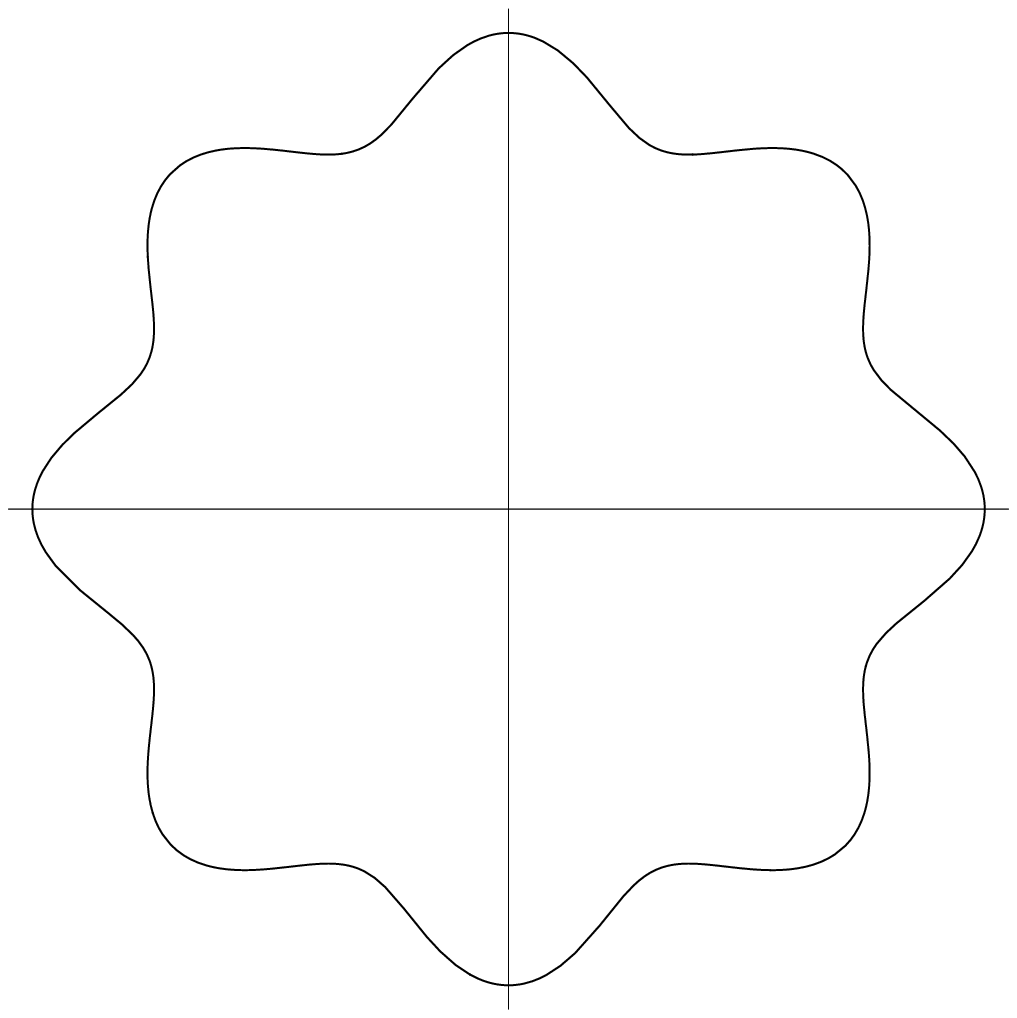}
\end{center}
\vskip -0.4 cm \caption{On the left NS5-branes distributed on a
circle of finite radius. They are deformed into the shape depicted
on the right in the same plane. The symmetry is broken from $SO(2)_A
\times SO(2)_B$ to $SO(2)_B$. The perturbation is driven by a chiral
primary operator whose spin is related to the number of modes as
$n=2(j+1)$.}
\end{figure}

\section{Turning on angular momentum with CFT operators }

The supergravity background corresponding to $k$ rotating NS5-branes was constructed in \cite{sferot}.
Quite a simplification occurs in the field-theory limit given by eqs. (14)-(17) of \cite{sferot}, where the
interested reader can find the explicit form of the metric, the antisymmetric tensor and the dilaton fields.
It was shown there that the Euclidean continuation of this background is obtained
as an $O(3,3)$ transformation
of the $SL(2,R)_{-k}/U(1) \times SU(2)_k$ CFT.

\no
For vanishing rotation parameter $a_i$, $i=1,2$, this
background becomes that in (\ref{jk9}). The CFT operators
driving it to the full solution are found by
expanding the worldsheet Lagrangian density in the asymptotic region and keeping the first correction.
For this we obtain the expression
\ba
\d \cL =  \left[(a_1^2-a_2^2) J^3_+ J_-^3   +  \del_+ t \del_- t
 +  (a_2-a_1)  \del_+ t J_-^3
 - (a_1+a_2) J_+^3 \del_- t \right] e^{-q\phi} + {\cal O}\left(e^{-2 q \phi}\right)\ ,
\label{perrtt}
\ea
where the time $t$ is represented by a timelike free boson and
\be
J^3_\pm =  \sin^2\th \del_\pm \phi_1 \pm \cos^2\th \del_\pm \phi_2\ ,
\ee
are chiral and antichiral (on shell) currents
of the $SU(2)_k$ WZW model (generated by $\s_3$).
The perturbation is clearly marginal and the dimension zero factor $e^{-q\phi}$
guarantees its normalizability.
We also note that the first term in (\ref{perrtt}) is also
responsible for spreading the NS5-branes from a point to a circle \cite{PreSfe} (see, fig.1).

\no
We conclude by mentioning that it remains to discuss systematically CFT deformations of the NS5 wold-volume.
In that respect appropriate supergravity solutions have been constructed in  \cite{Papadopoulos}.

 \end{document}